\documentclass[11pt,a4paper]{article}
\usepackage[margin=1in]{geometry}
\usepackage{amsmath,amssymb,amsfonts,amsthm}
\usepackage{physics}
\usepackage{bm}
\usepackage{graphicx}
\usepackage{caption}
\usepackage{subcaption}
\usepackage{authblk}
\usepackage{hyperref}
\usepackage[numbers,sort&compress]{natbib}
\usepackage{microtype}
\usepackage{mathtools}
\usepackage{siunitx}
\usepackage{enumitem}
\usepackage{float}

\title{Frequency Detuning and Interference-Induced Bohmian Chaos in a Two-Dimensional Anisotropic Harmonic Oscillator}

\author[1,3]{Umair Abdul Halim}
\author[2,3]{Nurisya Mohd Shah}
\author[2,3]{Chan Kar Tim}
\author[4]{Ahmad Hazazi Ahmad Sumadi}

\affil[1]{\small\textit {Centre of Foundation Studies in Science of Universiti Putra Malaysia (ASPutra), Universiti Putra Malaysia (UPM)\\
43400 UPM Serdang, Selangor Darul Ehsan, Malaysia}}

\affil[2]{\small\textit {Department of Physics, Faculty of Science, Universiti Putra Malaysia (UPM)\\
43400 UPM Serdang, Selangor Darul Ehsan, Malaysia}}

\affil[3]{\small\textit {Institute for Mathematical Research (INSPEM)\\
Universiti Putra Malaysia (UPM), 43400 UPM Serdang, Selangor Darul Ehsan, Malaysia}}

\affil[4]{\small\textit {Department of Applied Physics, Faculty of Science and Technology\\
Universiti Kebangsaan Malaysia (UKM), 43600 UKM Bangi, Selangor Darul Ehsan, Malaysia}}

\begin{document}
\maketitle

\begin{abstract}
We investigate the emergence of chaotic Bohmian trajectories in a three-mode superposition of the ground and first excited states of a two-dimensional anisotropic harmonic oscillator. The analysis focuses on the interference-induced phase structure of the wavefunction, which determines the Bohmian velocity field through its phase gradient. We show that the spatial extent of chaotic motion is controlled by the temporal coherence of the interference pattern, set by the detuning between oscillator modes. Near resonance, slow beating generates long-lived phase-gradient structures that repeatedly stretch and fold nearby trajectories, leading to more spatially extended chaotic regions. In contrast, strong detuning produces rapid temporal decorrelation of the phase field and confines chaotic dynamics to localized regions of configuration space. To quantify this behavior, we use a dimensionless coherence parameter comparing the beating time scale with a characteristic transport time. The results identify temporal coherence of the interference-induced phase field as a useful diagnostic for chaotic transport in low-dimensional Bohmian systems.
\end{abstract}

\section{Introduction}

Bohmian mechanics provides a trajectory-based formulation of quantum dynamics in which particles evolve deterministically under the guidance of the wavefunction \cite{Bohm1952a,Bohm1952b,Holland1993,BohmHiley1993,DurrTeufel2009}. In this framework, the wavefunction is written in polar form, the particle velocity is determined by the gradient of its phase, and the quantum potential enters through the quantum Hamilton--Jacobi equation. Beyond its foundational significance, Bohmian mechanics has also proven useful for analyzing quantum transport, tunneling, interference phenomena, and trajectory-based manifestations of chaos \cite{Wyatt2005,Benseny2014,deOliveira1998}.

A central question in this context is how chaotic motion can arise in a system governed by fully deterministic equations. In Bohmian mechanics, chaos does not originate from external randomness, but from the spatiotemporal structure of the guiding wavefunction itself. Even relatively simple superpositions of a few eigenstates can generate highly irregular trajectories, particularly when the wavefunction contains time-dependent interference patterns, quantum vortices, and rapidly varying phase gradients \cite{Frisk1997,Wisniacki2000,WisniackiPujals2005}. The emergence of chaos is therefore closely tied to the geometry and temporal evolution of the phase field.

Previous studies have shown that nodal structures, phase singularities, quantum vortices, and the global organization of the Bohmian velocity field play important roles in the emergence of chaotic trajectories \cite{Efthymiopoulos2007,ContopoulosEfthymiopoulos2008,Efthymiopoulos2009,ContopoulosTzemos2020}. Later work has further emphasized that chaotic Bohmian trajectories may arise from the broader phase structure of the wavefunction, even in situations where nodal dynamics alone do not fully determine the global behavior \cite{CesaMartinStruyve2016,TzemosContopoulos2022,TzemosContopoulos2023,TzemosContopoulos2024}. These results motivate an interpretation of Bohmian chaos in terms of the spatiotemporal organization of the phase field and its associated velocity gradients.

In parallel, many previous investigations of Bohmian trajectory complexity have employed incommensurate (irrational) frequency ratios, for which the phase evolution becomes quasi-periodic and densely explores phase space. While such settings can produce intricate trajectory patterns, the resulting complexity is often intertwined with dephasing associated with incommensurability. In the present work, we instead focus on rational and near-rational frequency ratios. This choice suppresses quasi-periodic phase filling and allows a clearer identification of chaos arising from interference-driven phase dynamics. Importantly, nodal structures and their associated singularities are independent of whether the frequency ratio is rational or irrational; however, restricting to rational ratios provides a cleaner setting in which interference effects can be isolated.

The system considered here is a minimal but nontrivial example: a three-mode superposition of the ground state and first excited states of a two-dimensional anisotropic harmonic oscillator. This model retains a simple analytic structure while exhibiting rich time-dependent interference. The key control parameter is the ratio of oscillator frequencies, which determines the beating frequency of the superposition and thus the temporal coherence of the interference pattern.

A central idea of this work is that the transition from localized to extended chaotic motion is governed by the competition between the beating time scale and the characteristic transport time of the Bohmian particle. Slow beating leads to long-lived phase coherence and persistent phase-gradient structures, resulting in sustained stretching and folding of trajectories. Fast beating, on the other hand, causes rapid temporal decorrelation and confines chaotic behavior to localized regions.

To quantify this competition, we introduce a dimensionless coherence parameter $\chi$ (defined in Sec.~\ref{sec:theory}) that measures the ratio between these two time scales. Large values of $\chi$ correspond to near-resonant conditions and coherent phase evolution, while small values correspond to strong detuning and rapid phase decorrelation. In this sense, the frequency ratio acts as a direct control parameter for the persistence of phase-gradient structures and the spatial extent of chaos.

Our numerical results indicate that chaotic behavior in this system is governed by the persistence or loss of coherence in the interference-induced phase-gradient field. In near-resonant regimes, slowly evolving interference produces long-lived phase-gradient structures that repeatedly stretch and fold Bohmian trajectories. In strongly detuned regimes, rapid temporal decorrelation suppresses sustained phase-gradient forcing, so chaotic behavior remains more localized. This provides a simple interference-based perspective on the transition between localized and extended Bohmian chaos.

The paper is organized as follows. Section~\ref{sec:theory} presents the theoretical framework and introduces the reduced amplitude and coherence parameter. Section~\ref{sec:numerics} describes the numerical procedure used to compute Bohmian trajectories, Lyapunov exponents, Poincar\'e sections, and phase-gradient diagnostics. Section~\ref{sec:results} analyzes Bohmian trajectories, velocity fields, and Lyapunov exponents for representative frequency ratios. Section~\ref{sec:discussion} interprets the results in terms of interference coherence and phase-gradient persistence. Section~\ref{sec:conclusion} summarizes the main findings.

\newpage

\section{Theoretical framework}
\label{sec:theory}

In Bohmian mechanics the wavefunction is written in polar form,
\begin{equation}
\psi(\mathbf r,t)=R(\mathbf r,t)\,e^{iS(\mathbf r,t)/\hbar},
\end{equation}
where $R\ge 0$ and $S$ are real. The particle velocity is determined by the phase gradient,
\begin{equation}
\dot{\mathbf r}=\frac{1}{m}\nabla S,
\label{eq:velocity}
\end{equation}
and the phase satisfies the quantum Hamilton--Jacobi equation
\begin{equation}
\partial_t S+\frac{(\nabla S)^2}{2m}+V+Q=0,
\qquad
Q(\mathbf r,t)=-\frac{\hbar^2}{2m}\frac{\nabla^2R}{R}.
\end{equation}
Thus, the Bohmian flow is governed by the spatiotemporal structure of the wavefunction through its amplitude and phase.

We consider a two-dimensional anisotropic harmonic oscillator,
\begin{equation}
H=\frac{p_x^2+p_y^2}{2m}+\frac{1}{2}m\left(\omega_x^2x^2+\omega_y^2y^2\right),
\end{equation}
and study the three-mode superposition of the ground state and the first excited states,
\begin{equation}
\psi(x,y,t)=
c_0\psi_{00}(x,y)e^{-iE_{00}t/\hbar}
+c_x\psi_{10}(x,y)e^{-iE_{10}t/\hbar}
+c_y\psi_{01}(x,y)e^{-iE_{01}t/\hbar},
\end{equation}
where \(c_0,c_x,c_y\in\mathbb C\) are complex expansion coefficients chosen so that the total wavefunction is normalized. For the oscillator eigenstates used here,
\begin{equation}
\psi_{00}=G(x,y),\qquad
\psi_{10}=\sqrt{2\alpha_x}\,x\,G(x,y),\qquad
\psi_{01}=\sqrt{2\alpha_y}\,y\,G(x,y),
\end{equation}
with
\begin{equation}
G(x,y)=
\frac{(\alpha_x\alpha_y)^{1/4}}{\sqrt{\pi}}
\exp\!\left[-\frac{1}{2}\left(\alpha_xx^2+\alpha_yy^2\right)\right],
\qquad
\alpha_{x,y}=\frac{m\omega_{x,y}}{\hbar},
\end{equation}
and energies
\begin{equation}
E_{00}=\hbar\left(\frac{\omega_x+\omega_y}{2}\right),\qquad
E_{10}=E_{00}+\hbar\omega_x,\qquad
E_{01}=E_{00}+\hbar\omega_y.
\end{equation}
Factoring out the common Gaussian envelope and the ground-state phase, the wavefunction can be written as
\begin{equation}
\psi(x,y,t)=G(x,y)e^{-iE_{00}t/\hbar}\Phi(x,y,t),
\qquad
\Phi(x,y,t)
=
c_0
+ c_x \sqrt{2\alpha_x}\, x\, e^{-i\omega_x t}
+ c_y \sqrt{2\alpha_y}\, y\, e^{-i\omega_y t}.
\label{eq:amplitude}
\end{equation}
The coefficients \(c_0,c_x,c_y\) contain the relative amplitudes and phases of the three components. The density therefore takes the form
\begin{equation}
|\psi(x,y,t)|^2=G^2(x,y)\,|\Phi(x,y,t)|^2,
\end{equation}
so that all interference effects are contained in $|\Phi|^2$. In particular, the overlap between the two first excited states produces
\begin{equation}
I_{10,01}(x,y,t)
=
4\sqrt{\alpha_x\alpha_y}\,G^2(x,y)\,x y\,
\Re\!\left[c_x c_y^{*} e^{-i(\omega_x-\omega_y)t}\right],
\label{eq:interference}
\end{equation}
which oscillates at the beating frequency
\begin{equation}
\Omega_b=|\omega_x-\omega_y|.
\end{equation}
This term is central because it controls the moving nodal pattern and the spatiotemporal modulation of the phase field.

The Bohmian velocity can be expressed in terms of the reduced amplitude $\Phi$,
\begin{equation}
\nabla S=\hbar\,\Im\!\left(\frac{\nabla\Phi}{\Phi}\right),
\end{equation}
so the velocity field is highly sensitive to the relative phases of the contributing modes. Near zeros of $\Phi$, the phase becomes ill-defined and its gradient exhibits large variations, producing strong velocity bursts and rapid local stretching of trajectories. The Gaussian factor $G$ does not determine the nodal geometry or the phase gradient, but it weights the density and contributes to the confining background through the quantum potential. Consequently, the central region, where the Gaussian envelope is largest and the polynomial factors $x$ and $y$ change sign most rapidly, plays a privileged role in the dynamics.

The transition from localized to extended chaotic motion is controlled by the temporal coherence of the interference pattern. To quantify this effect, we introduce the dimensionless coherence parameter
\begin{equation}
\chi=\frac{T_b}{T_{\mathrm{tr}}}
=\frac{2\pi v_0}{L\,|\omega_x-\omega_y|},
\label{eq:coherence}
\end{equation}
where \(T_b=2\pi/\Omega_b\) is the beating time associated with the detuning frequency \(\Omega_b=|\omega_x-\omega_y|\), and \(T_{\mathrm{tr}}\sim L/v_0\) is a characteristic transport time across the central interference region. In the present oscillator model, the length scale \(L\) is chosen as the characteristic width of the Gaussian-weighted central region. A natural estimate is obtained from the oscillator lengths
\begin{equation}
\ell_x=\sqrt{\frac{\hbar}{m\omega_x}},\qquad
\ell_y=\sqrt{\frac{\hbar}{m\omega_y}},
\end{equation}
so that one may take, for example,
\begin{equation}
L=(\ell_x\ell_y)^{1/2}.
\end{equation}
The velocity scale \(v_0\) is defined as a representative Bohmian speed in the same central region, for instance the root-mean-square value
\begin{equation}
v_0=
\left[
\left\langle
|\dot{\mathbf r}(x,y,t)|^2
\right\rangle_{\mathcal D}
\right]^{1/2},
\end{equation}
where \(\mathcal D\) denotes the central interference domain. Since the Bohmian velocity becomes formally singular at exact zeros of \(\Phi\), the quantity \(v_0\) should be understood as a representative finite velocity scale sampled away from exact nodal singularities. With these definitions, \(\chi\) compares the time scale over which the interference pattern reorganizes with the time scale over which a Bohmian trajectory samples the central interference region.

Large \(\chi\) corresponds to near resonance and slow beating, so that phase-gradient structures persist over times longer than the local transport time. This promotes sustained stretching and folding of nearby trajectories and favors extended chaotic motion. Small \(\chi\) corresponds to strong detuning, rapid temporal decorrelation of the phase pattern, and phase-gradient bursts that remain short-lived and spatially localized. In this framework, chaos is governed by interference coherence and frequency detuning through their control of the lifetime and spatial extent of phase-gradient structures.

This theoretical picture explains the numerical results reported below: the strongly detuned case supports localized chaotic channels coexisting with regular islands, whereas the near-resonant case sustains long-lived phase-gradient structures that support extended chaotic dynamics.

\newpage

\section{Numerical methods}
\label{sec:numerics}

All numerical calculations were carried out in dimensionless natural units. For convenience, we set
\[
m=\hbar=1.
\]
Because the system is anisotropic, the natural oscillator length scales in the two coordinate directions are
\[
\ell_x=\sqrt{\frac{\hbar}{m\omega_x}},
\qquad
\ell_y=\sqrt{\frac{\hbar}{m\omega_y}}.
\]
The corresponding momentum scales are
\[
p_x=\sqrt{m\hbar\omega_x},
\qquad
p_y=\sqrt{m\hbar\omega_y}.
\]
Time is measured in inverse oscillator-frequency units, and energies are measured in the corresponding oscillator energy scales \(\hbar\omega_x\) and \(\hbar\omega_y\).

The Bohmian equations of motion were integrated using the velocity field
\[
\dot{x}=v_x(x,y,t),\qquad \dot{y}=v_y(x,y,t),
\]
where
\[
v_x=\frac{\hbar}{m}\Im\left(\frac{\partial_x\Phi}{\Phi}\right),
\qquad
v_y=\frac{\hbar}{m}\Im\left(\frac{\partial_y\Phi}{\Phi}\right).
\]
For the three-mode state considered here,
\[
\partial_x\Phi=c_x\sqrt{2\alpha_x}e^{-i\omega_x t},
\qquad
\partial_y\Phi=c_y\sqrt{2\alpha_y}e^{-i\omega_y t}.
\]
Thus, the numerical integration uses the reduced amplitude \(\Phi(x,y,t)\) directly, while the real Gaussian envelope does not contribute to the spatial phase gradient.

The expansion coefficients used in the simulations were taken proportional to
\[
c_0=0.37-0.02i,\qquad
c_x=0.44+0.49i,\qquad
c_y=-0.49+0.44i,
\]
and the coefficient vector was normalized before numerical integration. The representative frequency ratios studied in detail were
\[
\omega_x:\omega_y=1{:}5
\]
and
\[
\omega_x:\omega_y=4{:}5,
\]
corresponding respectively to strongly detuned and near-resonant cases.

The Bohmian equations of motion were solved using Mathematica's \texttt{NDSolve} with adaptive time stepping. Therefore, no fixed integration time step was imposed. For the Lyapunov exponent calculation, the reference trajectory was integrated over the interval
\[
0\leq t\leq 7000,
\]
while the finite-time Lyapunov exponent was evaluated up to
\[
t_{\max}=2000 .
\]
The adaptive step-size procedure of \texttt{NDSolve} was used to control the numerical integration. In the Lyapunov calculation, the short-time integrations of nearby trajectories were performed with an increased maximum number of internal integration steps.

Near nodal points, where \(|\Phi|\) becomes small, the Bohmian velocity may become very large because the velocity field contains the ratio \(\nabla\Phi/\Phi\). In the numerical analysis, these near-nodal regions were monitored through the behavior of the trajectory, velocity, and phase-gradient magnitude. The resulting sharp velocity and phase-gradient fluctuations were interpreted as signatures of encounters with near-nodal regions, rather than as ordinary finite-amplitude velocity variations.

The finite-time Lyapunov exponent was computed by evolving two initially nearby Bohmian trajectories. The initial separation was chosen as
\[
\delta_0=10^{-3}.
\]
The separation between the two trajectories was periodically renormalized in order to avoid numerical saturation. The renormalization interval was
\[
\Delta t_{\mathrm{ren}}=0.1 .
\]
The finite-time Lyapunov exponent was estimated as
\[
\lambda(t)=\frac{1}{t}\sum_{k=1}^{n}
\ln\left(\frac{\delta_k}{\delta_0}\right),
\]
where \(\delta_k\) is the separation before the \(k\)-th renormalization. In the Lyapunov calculation, the representative initial condition was
\[
(x_0,y_0)=(2.0,0.0).
\]
A trajectory was classified as chaotic when the finite-time Lyapunov exponent approached a positive value over the integration interval.

Poincaré sections were constructed by stroboscopic sampling of Bohmian trajectories at the common period of the commensurate phase evolution. For the integer frequency ratios considered here, the sampling interval was
\[
T=2\pi,
\]
so that the sampled points were recorded at
\[
t_n=2\pi n,\qquad n=1,2,\ldots,N_{\mathrm{cyc}},
\]
where \(N_{\mathrm{cyc}}\) denotes the total number of stroboscopic cycles. In the simulations,
\[
N_{\mathrm{cyc}}=2500 .
\]
For each Poincaré section, \(600\) initial conditions were randomly sampled from the square
\[
(x_0,y_0)\in[-5,5]\times[-5,5].
\]
Each trajectory was integrated over the interval
\[
0\leq t\leq 2\pi N_{\mathrm{cyc}},
\]
and the corresponding stroboscopic points \((x(t_n),y(t_n))\) were plotted. The same sampling procedure was used when comparing the detuned and near-resonant cases.

\newpage

\section{Frequency Ratio as a Control Parameter}
\label{sec:results}

We now examine how the ratio of oscillator frequencies, \(\omega_y/\omega_x\), acts as a control parameter governing the transition between regular and chaotic Bohmian motion. Numerical simulations for several frequency ratios show that the qualitative structure of the trajectories depends sensitively on the detuning between the two excited modes. This dependence is expected from the theoretical framework above, since the detuning fixes the beating frequency \(\Omega_b=|\omega_x-\omega_y|\), and therefore controls the temporal coherence of the interference-induced phase-gradient field.

\begin{figure}[htbp]
  \centering
  \begin{subfigure}[t]{0.48\textwidth}
    \centering
    \includegraphics[width=\textwidth]{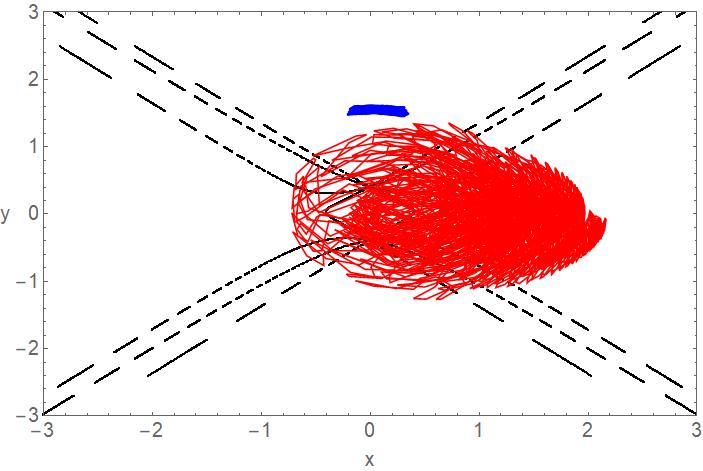}
   \caption{}
    \label{fig:traj_upper}
  \end{subfigure}
  \hfill
  \begin{subfigure}[t]{0.48\textwidth}
    \centering
    \includegraphics[width=\textwidth]{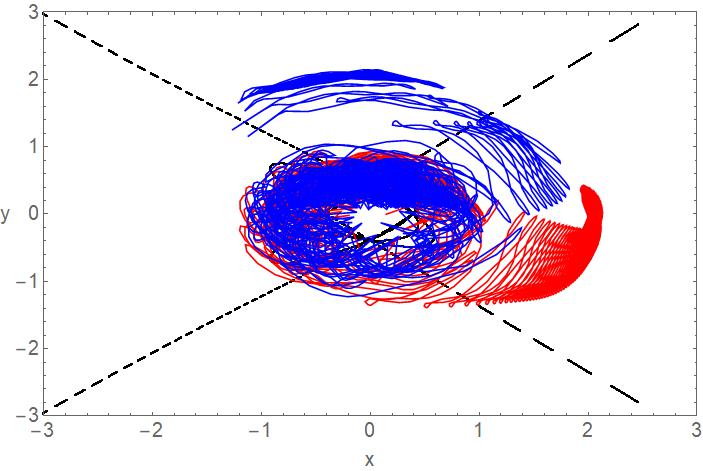}
    \caption{}
    \label{fig:traj_nodal}
  \end{subfigure}

 \caption{Comparison of Bohmian trajectories for two frequency ratios.
(a) For \(\omega_x:\omega_y=1{:}5\), regular and irregular trajectories coexist, with irregular motion mainly in the lateral regions.
(b) For \(\omega_x:\omega_y=4{:}5\), a broader set of trajectories is drawn into the central interference region and becomes irregular.}
  \label{fig:traj_14}
\end{figure}

For the strongly detuned case \(\omega_x:\omega_y=1{:}5\), trajectories starting in the upper or lower regions of configuration space display predominantly regular motion, while irregular behavior appears mainly for trajectories that pass close to regions of strong destructive interference. In contrast, trajectories initiated in the left and right regions tend to be guided toward the central interference region. Upon entering this region, the motion becomes highly erratic, reflecting repeated encounters with sharp variations in the phase field (Fig.~\ref{fig:traj_14}(a)). When the frequency ratio is changed to the near-resonant case \(\omega_x:\omega_y=4{:}5\), the qualitative mechanism remains similar, but the spatial extent of the irregular region increases significantly. In this case, trajectories from a much broader set of initial conditions are drawn into the central interference region and exhibit irregular motion (Fig.~\ref{fig:traj_14}(b)). This comparison indicates that, in the present three-mode system, smaller frequency detuning enhances the spatial extent of chaotic Bohmian dynamics.

\begin{figure}[htbp]
  \centering
  \begin{subfigure}[t]{0.48\textwidth}
    \centering
    \includegraphics[width=\textwidth]{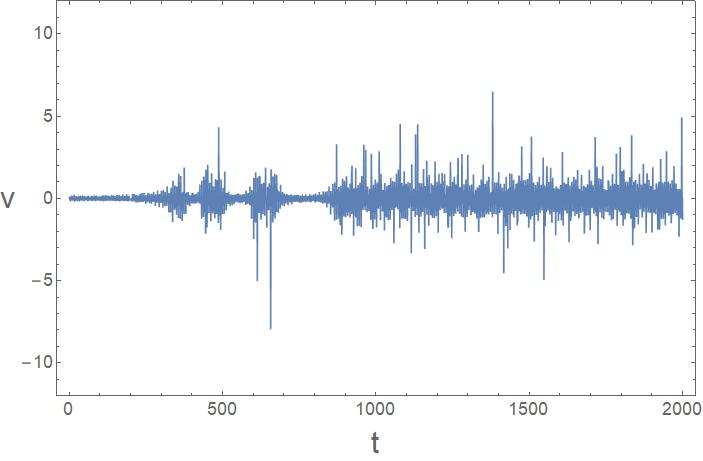}
   \caption{}
    \label{fig:velocity_15}
  \end{subfigure}
  \hfill
  \begin{subfigure}[t]{0.48\textwidth}
    \centering
    \includegraphics[width=\textwidth]{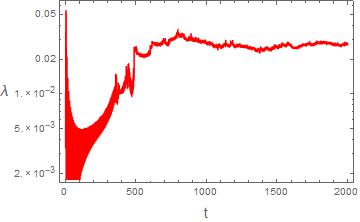}
    \caption{}
    \label{fig:lyapunov_15}
  \end{subfigure}

  \caption{
(a) Velocity evolution for the $\omega_x:\omega_y=1{:}5$ case with $(x_0,y_0)=(2.0,0.0)$, showing fluctuations as the trajectory enters the central interference region.
(b) The corresponding finite-time Lyapunov exponent $\lambda(t)$, which becomes positive after the transient phase, indicating chaotic Bohmian motion.
}
    \label{fig:velocity_lyapunov}
\end{figure}

The transition from regular to chaotic motion is also reflected in the time evolution of the Bohmian velocity. For the representative initial condition \((x_0,y_0)=(2.0,0.0)\) in the \(\omega_x:\omega_y=1{:}5\) case, the velocity initially exhibits smooth, nearly periodic oscillations, consistent with regular motion. As the trajectory approaches the central interference region, the velocity develops sharp fluctuations and loses its periodic character, indicating the onset of irregular dynamics (Fig.~\ref{fig:velocity_lyapunov}(a)). These fluctuations arise because the Bohmian velocity depends on \(\nabla\Phi/\Phi\), so it becomes highly sensitive to small values of \(|\Phi|\) and to rapid variations of the local phase.

This transition is quantitatively supported by the finite-time Lyapunov exponent. As shown in Fig.~\ref{fig:velocity_lyapunov}(b), the exponent initially fluctuates near zero during the regular phase, then rises and approaches a positive value after the transient regime. This positive finite-time value indicates exponential sensitivity to initial conditions over the integration interval and supports the identification of chaotic Bohmian motion. The agreement between the onset of irregular velocity fluctuations and the growth of the Lyapunov exponent provides a consistent trajectory-level signature of the regular-to-chaotic transition.

\newpage

\subsection{Interference-induced phase structure and the onset of Bohmian chaos}

The emergence of chaotic Bohmian trajectories in this three-mode superposition is rooted in how quantum interference shapes the phase field of the wavefunction and, through it, the velocity field. Since the particle velocity is determined by the phase gradient (Eq.~\eqref{eq:velocity}), rapid spatial variation or temporal instability of the phase $S(x,y,t)$ directly produces large and strongly fluctuating velocities. The phase structure therefore acts as the intermediary through which interference generates trajectory-level instability.

For the superposition studied here, all essential features of the dynamics are encoded in the reduced complex amplitude \(\Phi(x,y,t)\) introduced in Eq.~\eqref{eq:amplitude}. It contains the full time-dependent interference structure arising from the combination of the ground and first excited states. As a result, the probability density exhibits a nontrivial spatiotemporal pattern, with cross terms describing interference between the contributing modes. In particular, the overlap between the two first excited states generates an interference term proportional to
\(xy\,\Re[c_xc_y^*e^{-i(\omega_x-\omega_y)t}]\), which varies sinusoidally at the detuning frequency \(\Omega_b\). This contribution plays a central role in shaping the evolving nodal structure and the associated phase field.

The Bohmian velocity field is highly sensitive to this interference pattern through its dependence on $\Phi$. Because $\Phi(x,y,t)$ is a coherent superposition of contributions with different frequencies, both its magnitude and phase depend sensitively on the relative phases $(\omega_x t,\omega_y t)$. Near zeros of $\Phi$ (nodal points), the phase becomes singular and its gradient grows large, producing sharp velocity spikes and strong local stretching of trajectories.

The polynomial factor $xy$, together with the Gaussian envelope shown in Eq.~\eqref{eq:interference}, implies that interference effects are strongest near the origin, where the amplitude is maximal and the polynomial factors change sign most rapidly. Consequently, small spatial displacements in this region produce large relative variations in both amplitude and phase, making the phase gradient particularly sensitive to position and time. The central region therefore plays a privileged dynamical role for all frequency ratios.

The transition from localized to more spatially extended irregular motion is governed not by the mere presence of nodal points, but by the temporal organization of the interference pattern. The relevant control parameter is the beating frequency \(\Omega_b\), which sets the time scale over which the interference structure evolves. To quantify this effect, we use the coherence parameter \(\chi\) defined in Eq.~\eqref{eq:coherence}. This parameter compares the beating time \(T_b\) with the characteristic transport time \(T_{\mathrm{tr}}\) across the central interference region. The spatial scale \(L\) is estimated from the oscillator lengths, while \(v_0\) represents a root-mean-square Bohmian speed sampled in the central region. Since the Bohmian velocity is formally singular at exact zeros of \(\Phi\), this velocity scale is interpreted away from exact nodal singularities.

\begin{figure}[htbp]
\centering
\includegraphics[width=0.7\textwidth]{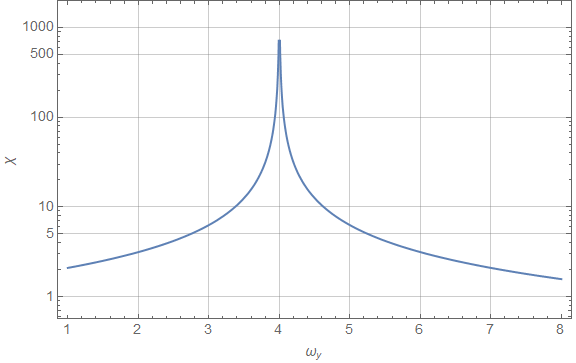}
\caption{
Coherence parameter \(\chi\) versus \(\omega_y\) for fixed \(\omega_x=4\), shown on a logarithmic scale. The divergence near \(\omega_y=\omega_x\) indicates vanishing detuning and enhanced phase coherence; the singular point is excluded.
}
\label{fig:chi_plot}
\end{figure}

The dependence of \(\chi\) on \(\omega_y\) is shown in Fig.~\ref{fig:chi_plot}. A pronounced divergence occurs as \(\omega_y \to \omega_x\), corresponding to vanishing detuning and long-lived phase coherence. Consistent with this trend, the strongly detuned trajectory case \(\omega_x:\omega_y=1{:}5\) corresponds to a small-\(\chi\) regime, whereas the near-resonant case \(\omega_x:\omega_y=4{:}5\) lies much closer to the large-\(\chi\) regime. Thus, for large \(\chi\), the interference pattern evolves slowly compared with the transport time, allowing phase-gradient structures to influence trajectories over longer intervals and favoring chaotic motion over a broader region of configuration space. In contrast, for small \(\chi\), rapid temporal decorrelation suppresses sustained interaction with these structures, so chaos remains localized or transient. The symmetry of the curve about \(\omega_y=\omega_x\) indicates that the dynamics depend primarily on the magnitude of the detuning. Overall, this supports the interpretation that the transition to chaos is governed by a competition between interference beating and Bohmian transport, establishing \(\chi\) as a useful diagnostic for interference-induced chaotic dynamics.

\begin{figure}[htbp]
  \centering
  \begin{subfigure}[t]{0.48\textwidth}
    \centering
    \includegraphics[width=\textwidth]{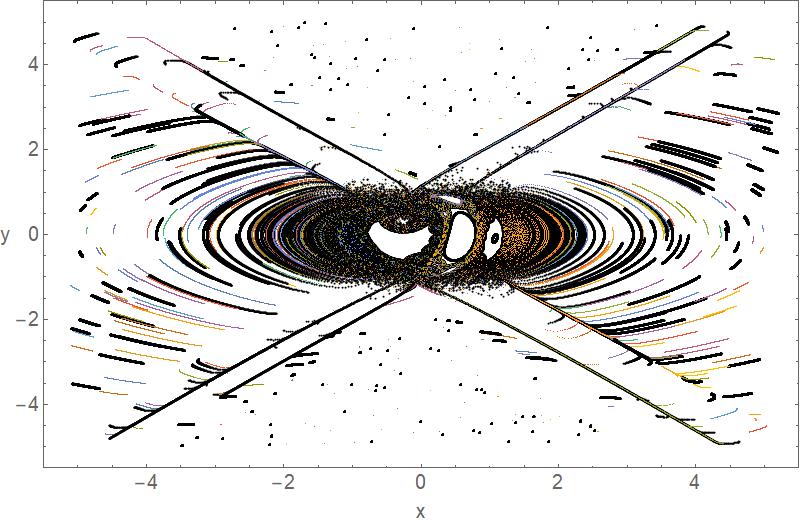}
   \caption{}
    \label{fig:poincare_15}
  \end{subfigure}
  \hfill
  \begin{subfigure}[t]{0.48\textwidth}
    \centering
    \includegraphics[width=\textwidth]{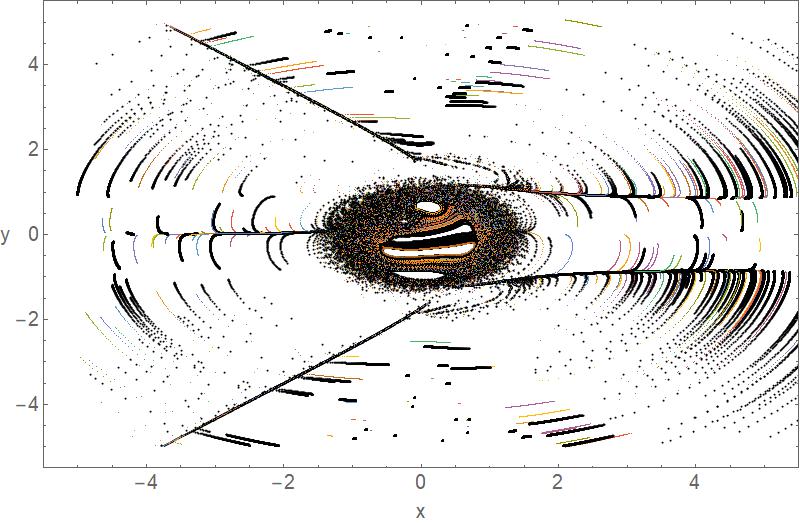}
    \caption{}
    \label{fig:poincare_45}
  \end{subfigure}

  \caption{Poincaré sections for two frequency ratios.
(a) For $\omega_x:\omega_y=1{:}5$, regular motion dominates, with irregularity mainly near lateral channels.
(b) For $\omega_x:\omega_y=4{:}5$, the irregular region expands and more trajectories approach the central interference region.}
  \label{fig:Poincare_combined}
\end{figure}

These mechanisms manifest clearly in the Poincaré sections shown in Fig.~\ref{fig:Poincare_combined}, which provide a global phase--space representation of the local interference-driven dynamics. For the strongly detuned ratio $\omega_x:\omega_y=1{:}5$ (Fig.~\ref{fig:Poincare_combined}(a)), the Poincaré map exhibits a pronounced coexistence of regular and chaotic motion. Regular trajectories dominate the upper and lower regions of configuration space, appearing as smooth invariant curves that persist over long times.

This behavior reflects the large beating frequency $\Omega_b$, which induces rapid temporal oscillations of the interference phase. As a result, nodal points undergo fast oscillatory motion and the phase-gradient field fluctuates on short time scales, leading to rapid temporal decorrelation. This suppresses sustained trajectory instability over extended regions of configuration space. Nevertheless, the resulting chaos is anisotropic and appears predominantly in the left and right channels connecting to the central interference region. In these lateral lobes (large $|x|$, small $|y|$), the dynamics are effectively governed by the slowly oscillating $x$-mode, while the rapidly oscillating $y$-mode becomes dynamically negligible due to its suppressed amplitude. As a result, the local interference structure is well approximated by a reduced two-mode superposition dominated by the slower frequency, allowing coherent phase gradients to persist along these channels. Consequently, the wavefunction reduces effectively to
\begin{equation}
\Phi(x,y,t) \approx c_0+\tilde c_x x e^{-i\omega_x t},
\qquad
\tilde c_x=c_x\sqrt{2\alpha_x}.
\end{equation}

Trajectories entering these regions are gradually guided toward the central interference zone, where they encounter repeated velocity bursts associated with sharp variations in the phase field. This leads to localized chaotic motion confined to the channel regions, rather than spatially extended chaotic transport.

A qualitatively different situation arises for the near-resonant ratio $\omega_x:\omega_y=4{:}5$, where $\Omega_b$ is small. The slow beating leads to long-lived phase coherence in the interference pattern. Nodal points move more smoothly and persist over extended time intervals, generating phase-gradient structures that remain dynamically influential across large portions of configuration space. Because no strong time-scale separation exists between the $x$- and $y$-modes, coherent stretching is no longer confined to specific directions. The corresponding Poincar\'e section in Fig.~\ref{fig:Poincare_combined}(b) therefore exhibits a much denser distribution of points over a broader region, consistent with the emergence of spatially extended irregular dynamics.

\begin{figure}[htbp]
  \centering

  \begin{subfigure}[t]{0.48\textwidth}
    \centering
    \includegraphics[width=\textwidth]{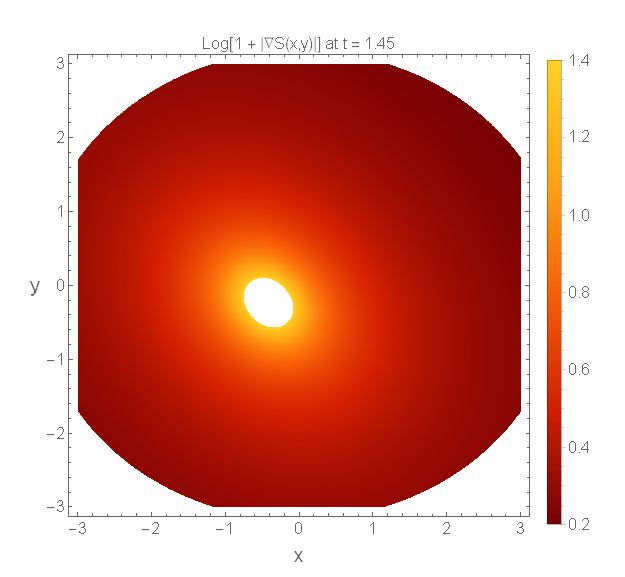}
    \caption{}
    \label{fig:phase_gradient_map}
  \end{subfigure}
  \hfill
   \begin{subfigure}[t]{0.48\textwidth}
    \centering
    \includegraphics[width=\textwidth]{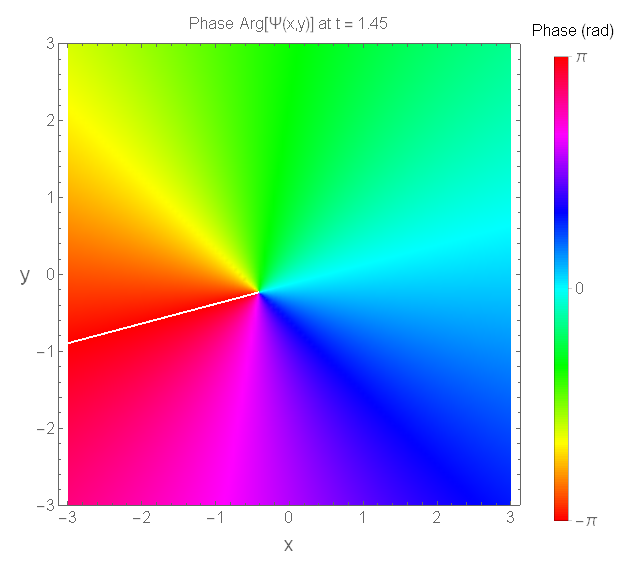}
   \caption{}
    \label{fig:phase_map}
  \end{subfigure}

\caption{
Spatial phase-gradient magnitude and phase structure at a representative time.
(a) $\log(1+|\nabla S|)$ highlights high-gradient regions driving strong Bohmian velocities.
(b) The phase field $S(x,y)=\arg\Psi(x,y)$ shows phase variation modulo $2\pi$, with abrupt transitions marking branch cuts.
}
\label{fig:phase_grad}
\end{figure}

These dynamical differences are further illustrated by the phase gradient and phase maps shown in Fig.~\ref{fig:phase_grad}. Panel (a) displays the phase gradient magnitude, while panel (b) shows the corresponding phase field. The phase gradient map highlights regions where the magnitude $|\nabla S|$ is large. In the near resonant $4{:}5$ case, these high gradient regions persist and drift slowly across the central region, reflecting the coherent evolution of nodal structures. This behavior is accompanied by a slowly evolving branch cut, visible as abrupt color transitions in the phase plot, which traces regions of rapid phase variation and enhanced gradients in the phase field. In contrast, for the detuned $1{:}5$ case, regions of large $|\nabla S|$ are short-lived and spatially localized. Rapid beating leads to fast temporal decorrelation of the phase pattern, causing the branch cuts to sweep quickly across the domain. Consequently, large gradients arise only transiently, producing brief velocity bursts that confine irregular motion to the central region and lateral channels.

\begin{figure}[H]
\centering
\begin{subfigure}{0.48\textwidth}
\centering
\includegraphics[width=\linewidth]{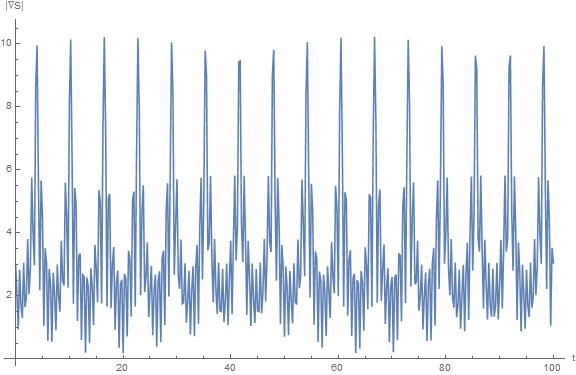}
\caption{$\omega_x:\omega_y = 1{:}5$}
\end{subfigure}
\hfill
\begin{subfigure}{0.48\textwidth}
\centering
\includegraphics[width=\linewidth]{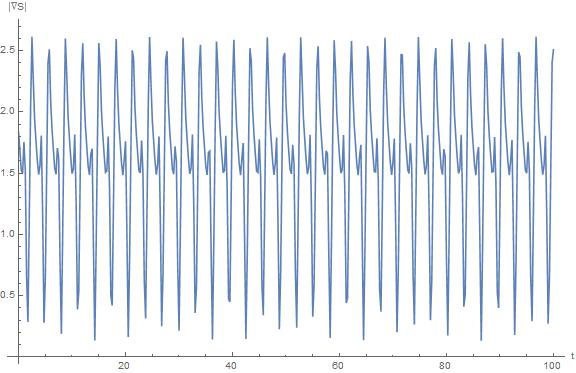}
\caption{$\omega_x:\omega_y = 4{:}5$}
\end{subfigure}

\caption{
Temporal evolution of \( |\nabla S(x_0,y_0,t)| \) at a fixed probe point for two frequency ratios.
(a) In the detuned \(1{:}5\) case, repeated spikes indicate rapid nodal motion and short temporal coherence.
(b) In the near-resonant \(4{:}5\) case, more regularly modulated and temporally correlated oscillations indicate slower nodal evolution and persistent phase-gradient fluctuations.
}

\label{fig:phase_grad_compare}
\end{figure}

This is consistent with the temporal evolution of the phase-gradient magnitude 
\(|\nabla S(x_0,y_0,t)|\) at a fixed probe point 
(Fig.~\ref{fig:phase_grad_compare}). For the detuned ratio 
\(\omega_x:\omega_y=1{:}5\) 
(Fig.~\ref{fig:phase_grad_compare}(a)), the signal contains repeated 
high-amplitude spikes, reflecting rapid nodal motion and short temporal 
coherence of the velocity field. These spikes correspond to brief encounters 
with near-nodal regions where the wavefunction amplitude becomes small. 
For the near-resonant ratio \(\omega_x:\omega_y=4{:}5\) 
(Fig.~\ref{fig:phase_grad_compare}(b)), the phase-gradient time series becomes 
more regularly modulated and temporally correlated, indicating slower nodal 
evolution and a more persistent influence of the phase-gradient field on the 
Bohmian trajectory. Together, the spatial maps and time series support the 
interpretation that strong detuning produces localized, spike-dominated 
instability, whereas near resonance produces more spatially extended irregular 
dynamics.

\section{Discussion}
\label{sec:discussion}

Our results show that the onset of Bohmian chaos in this three-mode superposition is governed primarily by the interference structure of the wavefunction and its associated phase dynamics. The frequency ratio acts as a spectral control parameter that determines how coherently the contributing modes interfere in time. In near-resonant cases, the interference pattern evolves slowly, producing longer-lived phase-gradient structures near the Gaussian-weighted center. In strongly detuned cases, the interference pattern decorrelates more rapidly, and the resulting phase-gradient fluctuations remain spatially localized and short-lived.

A central feature of the present system is the concentration of chaotic behavior near the origin of configuration space. This does not indicate the action of a direct central force. Rather, it reflects the fact that the superposition of the ground state and the first two excited states generates the strongest overlap near the center, where the Gaussian envelope is largest and where the polynomial factors change most rapidly. As a result, small changes in the relative phases of the contributing modes can produce large variations in the local interference pattern, which in turn create strong gradients in the phase field.

In Bohmian mechanics, the velocity field is determined by the phase through
\begin{equation}
\dot{\mathbf{r}}=\frac{1}{m}\nabla S,
\end{equation}
so the relevant dynamical quantity is not a direct attractive force, but the spatial and temporal structure of $S(x,y,t)$. The quantum potential is therefore best viewed as an indirect mediator of the dynamics, since it is fully determined by the wavefunction and influences the evolution only through its effect on the phase structure. In particular, rapid spatial and temporal changes in the interference pattern reshape the phase field, generate large values of $|\nabla S|$, and produce strong velocity bursts. These bursts stretch and fold nearby trajectories, leading to sensitive dependence on initial conditions and chaotic Bohmian motion.

The role of the detuning frequency
\begin{equation}
\Omega_b = |\omega_x-\omega_y|
\end{equation}
is especially important. It sets the time scale over which the interference pattern reorganizes. To quantify this effect, we introduced the dimensionless coherence parameter
\begin{equation}
\chi = \frac{2\pi v_0}{L\,|\omega_x-\omega_y|},
\end{equation}
which measures the ratio between the beating time scale and the characteristic transport time of the particle. Large values of $\chi$ correspond to slow beating and long-lived phase coherence, while small values of $\chi$ correspond to rapid beating and fast temporal decorrelation. This provides a compact diagnostic for distinguishing localized irregular motion from more spatially extended chaotic transport.

This interpretation is consistent with the Poincar\'e sections. For the strongly detuned case $\omega_x:\omega_y=1{:}5$, regular islands survive in the upper and lower regions, while chaotic motion is confined mainly to the left and right channels leading toward the center. In the near-resonant case $\omega_x:\omega_y=4{:}5$, the corresponding Poincar\'e section becomes much denser over a broader region, consistent with the expansion of irregular dynamics across configuration space.

The phase and phase-gradient plots support the same interpretation from a local viewpoint. In the near-resonant case, the high-gradient structures evolve more slowly and remain dynamically influential for longer times, whereas in the strongly detuned case they appear mainly as short-lived fluctuations. The time series of $|\nabla S(x_0,y_0,t)|$ further supports this distinction, showing spike-dominated behavior for $\omega_x:\omega_y=1{:}5$ and more temporally correlated modulation for $\omega_x:\omega_y=4{:}5$.

Several limitations should be noted. First, the coherence parameter \(\chi\) is introduced here as a diagnostic for the present three-mode oscillator model, and its applicability to more general Bohmian systems remains to be established. Second, the classification of regular and chaotic regions is based on finite-time Lyapunov exponents and therefore depends on the integration time, the sampling of initial conditions, and the numerical thresholds employed. Third, although the present analysis emphasizes the persistence of interference-induced phase-gradient structures, it does not attempt to provide a complete classification of all possible local mechanisms of Bohmian chaos. In this sense, the present perspective is best understood as complementary to other analyses of nodal structures, phase singularities, and local flow geometry. Rather than replacing such descriptions, the present results suggest that the temporal coherence of the interference-induced phase-gradient field provides an additional organizing principle governing whether chaotic stretching remains spatially localized or extends across larger regions of configuration space.

\section{Conclusion}
\label{sec:conclusion}

Within the present three-mode anisotropic oscillator, interference coherence appears to govern the transition from localized to more spatially extended chaotic transport. The key control parameter is the frequency detuning, which regulates the temporal coherence of the interference pattern and the persistence of large phase gradients. Near resonance, long-lived gradient structures produce sustained velocity instabilities and more spatially extended chaotic motion, whereas strong detuning confines instability to more localized regions.

Our results indicate that chaos in this system is governed by the spatiotemporal organization of the interference-induced phase field. The coherence parameter \(\chi\) offers a simple and physically transparent measure of this effect, linking the beating time scale to the transport time scale. Within the present three-mode oscillator model, interference coherence therefore provides a useful diagnostic for the transition from localized to more spatially extended chaotic transport. Whether the same diagnostic applies to broader classes of Bohmian systems remains an interesting direction for future work.

\section*{Acknowledgments}
This work is supported by Geran Putra - Inisiatif Putra Muda (GP-IPM) Universiti Putra Malaysia Grant (GP-IPM/2024/9790000).


\begin{thebibliography}{99}

\bibitem{Bohm1952a}
D.~Bohm,
``A Suggested Interpretation of the Quantum Theory in Terms of `Hidden' Variables. I,''
Phys. Rev. \textbf{85}, 166--179 (1952).

\bibitem{Bohm1952b}
D.~Bohm,
``A Suggested Interpretation of the Quantum Theory in Terms of `Hidden' Variables. II,''
Phys. Rev. \textbf{85}, 180--193 (1952).

\bibitem{Holland1993}
P.~R.~Holland,
\textit{The Quantum Theory of Motion: An Account of the de Broglie--Bohm Causal Interpretation of Quantum Mechanics}
(Cambridge University Press, Cambridge, 1993).

\bibitem{Wyatt2005}
R.~E.~Wyatt,
\textit{Quantum Dynamics with Trajectories: Introduction to Quantum Hydrodynamics}
(Springer, New York, 2005).

\bibitem{Benseny2014}
A.~Benseny, G.~Albareda, A.~S.~Sanz, J.~Mompart, and X.~Oriols,
``Applied Bohmian mechanics,''
Eur. Phys. J. D \textbf{68}, 286 (2014).

\bibitem{Frisk1997}
H.~Frisk,
``Properties of the trajectories in Bohmian mechanics,''
Phys. Lett. A \textbf{227}, 139--142 (1997).

\bibitem{WisniackiPujals2005}
D.~A.~Wisniacki and E.~R.~Pujals,
``Motion of vortices implies chaos in Bohmian mechanics,''
Europhys. Lett. \textbf{71}, 159--165 (2005).

\bibitem{Efthymiopoulos2007}
C.~Efthymiopoulos, C.~Kalapotharakos, and G.~Contopoulos,
``Nodal points and the transition from ordered to chaotic Bohmian trajectories,''
J. Phys. A: Math. Theor. \textbf{40}, 12945--12972 (2007).

\bibitem{ContopoulosEfthymiopoulos2008}
G.~Contopoulos and C.~Efthymiopoulos,
``Ordered and chaotic Bohmian trajectories,''
Celest. Mech. Dyn. Astron. \textbf{102}, 219--239 (2008).

\bibitem{Efthymiopoulos2009}
C.~Efthymiopoulos, C.~Kalapotharakos, and G.~Contopoulos,
``Origin of chaos near critical points of quantum flow,''
Phys. Rev. E \textbf{79}, 036203 (2009).

\bibitem{ContopoulosTzemos2020}
G.~Contopoulos and A.~C.~Tzemos,
``Chaos in Bohmian Quantum Mechanics: A short review,''
Regul. Chaotic Dyn. \textbf{25}, 476--495 (2020).

\bibitem{CesaMartinStruyve2016}
A.~Cesa, J.~Martin, and W.~Struyve,
``Chaotic Bohmian trajectories for stationary states,''
J. Phys. A: Math. Theor. \textbf{49}, 395301 (2016).

\bibitem{TzemosContopoulos2022}
A.~C.~Tzemos and G.~Contopoulos,
``Bohmian quantum potential and chaos,''
Chaos, Solitons \& Fractals \textbf{160}, 112151 (2022).

\bibitem{TzemosContopoulos2023}
A.~C.~Tzemos and G.~Contopoulos,
``Unstable points, ergodicity and Born's rule in 2D Bohmian systems,''
Entropy \textbf{25}, 1089 (2023).

\bibitem{TzemosContopoulos2024}
A.~C.~Tzemos and G.~Contopoulos,
``A comparison between classical and Bohmian quantum chaos,''
Chaos, Solitons \& Fractals \textbf{188}, 115524 (2024).


\bibitem{DurrTeufel2009}
D.~D\"urr and S.~Teufel,
\textit{Bohmian Mechanics: The Physics and Mathematics of Quantum Theory}
(Springer, Berlin, 2009).

\bibitem{BohmHiley1993}
D.~Bohm and B.~J.~Hiley,
\textit{The Undivided Universe: An Ontological Interpretation of Quantum Theory}
(Routledge, London, 1993).

\bibitem{deOliveira1998}
H.~P.~de Oliveira, D.~Jonathan, and J.~Ellis,
``Bohmian trajectories and quantum chaos,''
Physica A \textbf{261}, 534--545 (1998).

\bibitem{Wisniacki2000}
D.~A.~Wisniacki,
``Quantum vortices and trajectory complexity,''
Phys. Rev. E \textbf{63}, 016615 (2000).

\end{thebibliography}
\end{document}